\documentclass[preprintnumbers,article,amsmath,amssymb,floatfix,10pt,prd,superscriptaddress,nofootinbib]{revtex4}

\usepackage{doi}
\usepackage{hyperref}
\hypersetup{
  colorlinks=true,        
  linkcolor=magenta,         
  citecolor=red,         
}

\usepackage{graphicx}
\usepackage{dcolumn}
\usepackage{bm}
\usepackage{color}
\usepackage{enumitem}
\usepackage{amsmath}
\usepackage{amssymb}

\newcommand{\beq}{\begin{equation}}
\newcommand{\eeq}{\end{equation}}
\newcommand{\bea}{\begin{eqnarray}}
\newcommand{\eea}{\end{eqnarray}}

\usepackage{bbm}
\usepackage{amsfonts}
\usepackage{mathrsfs}
\usepackage{latexsym}
\usepackage{epsfig}
\usepackage{epstopdf}
\usepackage{epstopdf}
\usepackage{graphicx}
\usepackage{amssymb}
\usepackage{amsmath}
\usepackage{dcolumn}
\usepackage{bm}
\usepackage{color}
\usepackage{comment}
\usepackage{xcolor}
\usepackage{xcolor}
\usepackage{orcidlink}

\begin{document}
\title{Generalized Catability of Relativistic Quantum States Measurement\\ in a Unified Lie-Algebraic Foldy-Wouthuysen (FW) Framework}

\author{Abdelmalek Bouzenada\orcidlink{0000-0002-3363-980X}}
\email{abdelmalekbouzenada@gmail.com}
\affiliation{Laboratory of Theoretical and Applied Physics, Echahid Cheikh Larbi Tebessi University, 12001, Algeria}
\affiliation{Research Center of Astrophysics and Cosmology, Khazar University, Baku AZ1096, Azerbaijan}

\author{Rami Ahmad El-Nabulsi\orcidlink{0000-0002-3363-980X}}
\email{el-nabulsi@cesnet.cz (corresp. author)}
\affiliation{Center of Excellence in Quantum Technology, Faculty of Engineering, Chiang Mai University, Chiang Mai 50200, Thailand}
\affiliation{Quantum-Atom Optics Laboratory and Research Center for Quantum Technology, Faculty of Science, Chiang Mai University, Chiang Mai 50200, Thailand}
\affiliation{Department of Optical Networks CESNET a. l. e Generala Pıky 430/26, Prague, Czech Republic}

\begin{abstract}
In this work, a unified Lie-algebraic formulation of catability is constructed for relativistic quantum systems with arbitrary spin within this framework. In this case, the analysis starts with constructing catability as a quantitative measure for superposed coherent states, where coherence structure and quantum interference properties are studied using algebraic representations in this framework. Also, a generalized Foldy-Wouthuysen transformation is formulated within a Lie algebraic framework, delivering a systematic procedure for block-diagonalization of relativistic Hamiltonians and separation of positive- and negative-energy components in this framework. Within this formalism, a phase-sensitive catability operator is introduced to study phase correlations and coherence effects in the relativistic quantum dynamics framework. The approach is applied to Dirac spin-$1/2$ particles, where relativistic fermionic catability is analyzed in relation to spinorial structures and symmetry generators framework. The formalism is extended through a unified geometric and Lie-algebraic treatment, establishing a consistent description of catability in a relativistic quantum mechanics framework. In this context, the generalized framework is constructed for arbitrary spin-$s$ fields, enabling investigation of higher-spin relativistic quantum states within the same algebraic structure framework. In this context, the obtained results show a generalized theoretical platform for investigating relativistic quantum coherence, superposition effects, and algebraic symmetries in the framework of fermionic and bosonic systems.
\end{abstract}

\maketitle

\date{\today}


\section{Introduction}

Lie algebraic techniques constitute a rigorous mathematical framework for treating nonlinear differential equations appearing in general relativity, relativistic astrophysics, and cosmological models. In this case, the approach based on continuous transformation groups introduced by Lie establishes a systematic procedure for identifying invariant structures of differential equations and for deriving exact analytical solutions through symmetry reduction methods \cite{Olver1998, Ovsiannikov1982}. Within this formulation, Lie point symmetries correspond to infinitesimal transformations that map a solution space of a differential equation onto itself while maintaining the form of the governing equations. Also, these symmetry generators enable the reduction of partial differential equations to ordinary differential equations through the construction of similarity variables and invariant ansätze. Also, the associated algebraic structure is characterized by Lie algebras whose generators satisfy defined commutation relations that encode the invariance properties of the system \cite{Basarab2001}. The classification of differential equations according to their Lie algebraic structure provides an efficient framework for identifying integrable models and for constructing exact relativistic solutions. In addition, symmetry-preserving numerical algorithms and invariant discretization schemes have been developed in mathematical physics to maintain geometric consistency at the discrete level \cite{Bourlioux2006}. In relativistic astrophysics, Lie symmetry methods have been applied to the Einstein field equations describing compact objects, radiating configurations, and anisotropic matter distributions. Various studies have shown that symmetry generators reduce the nonlinear structure of the gravitational field equations and yield exact families of physically consistent stellar models \cite{Abebe2014a, Abebe2014b, Abebe2014c}. In particular, Lie group analysis has been used to obtain invariant solutions for radiating stars and conformally flat geometries. Moreover, the application of Lie symmetries to embedding equations for shear-free and conformally flat spherical spacetimes was investigated by Paliathanasis et al. \cite{Paliathanasis2022}, where invariant transformations were employed to simplify the embedding conditions and construct exact analytical solutions. In this framework, the Karmarkar condition admits Lie point symmetries under spherical symmetry, allowing coordinate transformations that reduce the complexity of the embedding relations. Since the embedding condition is purely geometric, the resulting exact solutions are independent of the underlying gravitational theory and can be applied in both Einstein gravity and modified gravity models. Furthermore, when appropriate junction conditions at the stellar boundary are imposed, these solutions can be used to construct relativistic compact star models and cosmological configurations with physically consistent behavior.

The Dirac equation (DE) is a central formulation of relativistic quantum mechanics, providing a consistent framework that combines quantum mechanics with special relativity and establishes the foundation of the relativistic quantum field theory framework \cite{D1} in a theoretical physics context. This formalism has produced significant developments in multiple branches of theoretical physics, including quantum electrodynamics \cite{D2} and relativistic particle dynamics investigation studies and additional analysis \cite{D3, D4}. Within the nuclear physics framework, the DE has been extensively used for analysis of intrinsic nucleonic properties \cite{D5}, while extensions include superheavy nuclei \cite{D6}, correlated quantum many-body systems \cite{D7}, chiral-symmetric quantum chromodynamics \cite{D8}, and antiparticle contributions in relativistic interactions for additional detailed analysis \cite{D9}. Further investigations based on the DE include electron vortex states \cite{D10}, coherent quantum manipulation methods \cite{D11}, and modern relativistic formulations with analytical approaches within a relativistic analytical framework study extended analysis \cite{D12, D13, D14, D15, D16}. A key aspect of the DE is the appearance of spin and pseudo-spin symmetries in the structure of the Dirac Hamiltonian formulation within the relativistic quantum mechanical Hamiltonian framework structure analysis \cite{D17, D18, D19, D20, D21, D22, D23, D24, D25, D26}. Spin symmetry corresponds to nearly degenerate doublets defined by quantum numbers $(n,l,j=l\pm s)$, describing antinucleon spectra together with spin-orbit coupling effects in hadronic systems within a relativistic nuclear interaction theoretical framework analysis consistent description \cite{D20, D21, D22, D23, D24}. In contrast, pseudo-spin symmetry corresponds to quasi-degenerate configurations between $(n,l,j=l+1/2)$ and $(n-1,l+2,j=l+3/2)$ states \cite{BZ11, BZ12, BZ13, BZ14, BZ15, BZ16, BZ17, BZ18, BZ19, BZ20, BZ21}, and it plays a substantial role in the analysis of nuclear deformation \cite{D27, D28}, identical rotational bands \cite{D29, D30, D31}, magnetic moment distributions \cite{D32, D33, BZ1, BZ2, BZ3, BZ4, BZ5, BZ6, BZ7, BZ8, BZ9, BZ10}, modifications of nuclear magic numbers \cite{D34, D35, D36}, and effective shell-model configurations within a relativistic nuclear structure analysis framework, including collective motion effects and effective interaction-based shell model descriptions and symmetry considerations aspects \cite{D37}.

Coherent cat states originate from the quantum superposition of quasi-classical wave packets characterized by macroscopically separated phases, a structure first motivated by Schrödinger’s analysis of quantum superposition \cite{SS1}. Their nonclassical features arise from interference structures in the Wigner phase-space representation, which cannot be reproduced within classical probability theory and are highly sensitive to external perturbations. In this framework, these states serve as a controlled model for decoherence processes, coherence attenuation, and the quantum-to-classical transition, while their pronounced phase dependence allows precision enhancement in metrological protocols beyond the standard quantum limit \cite{SS12, SS13}. From an experimental perspective, coherent cat states have been realized and coherently controlled in optical systems, superconducting cavities, trapped-ion platforms, and hybrid architectures, confirming their applicability in quantum information processing \cite{SS17, SS18, SS19, SS20, SS21, SS22, SS23, SS24, SS25, SS26, SS27, Ref1, Ref2, Ref3}. Nevertheless, the preservation of coherence at large excitation numbers and within scalable quantum networks remains restricted by decoherence mechanisms and control imperfections \cite{SS2, SS3, SS4, SS5, SS6, SS7, SS8, SS9, SS10, SS11}. To overcome these limitations, methods such as state breeding, concatenated encoding schemes, and bosonic error-correction protocols have been developed to enhance stability and increase resilience against noise \cite{SS14, SS15, SS16}. As a result, coherent cat states represent a relevant platform for analyzing quantum interference effects and simultaneously function as a resource for fault-tolerant computation, precision metrology, and quantum sensing applications \cite{SS8, SS14, SS16, SS28, SS29, BZ22, BZ23, BZ2, BZ25, BZ26, BZ27, BZ28, BZ29, BZ30, BZ31, BZ32}.

The structure of this paper follows. In Sec. (\ref{S2}), the concept of catability is introduced as a quantitative metric for analyzing superposed coherent states, focusing on its mathematical structure and physical interpretation in the quantum systems framework. In this case, Section (\ref{S3}) develops the generalized Foldy-Wouthuysen transformation within a Lie algebraic framework, establishing the algebraic structure necessary for diagonalizing relativistic Hamiltonians and separating positive- and negative-energy sectors in a quantum theory framework. Also, in Sec. (\ref{S4}), a phase-sensitive catability operator is constructed via Lie-algebraic representations, enabling analysis of the interference and phase-dependent quantum coherence properties structure. Section (\ref{S5}) extends the formalism to relativistic fermionic systems by investigating catability in Dirac spin-$1/2$ quantum dynamics, where spinorial degrees of freedom and relativistic corrections are examined in detail. In Sec. (\ref{S6}), the Dirac spin-$1/2$ description is further generalized within a unified Lie-algebraic and geometric framework, establishing a connection between algebraic symmetries and geometric structures of relativistic quantum states. Section (\ref{S7}) presents a generalized formulation of catability for arbitrary spin-$s$ fields, extending the unified Lie algebraic treatment beyond fermionic systems to higher-spin quantum fields. In this context, our results and physical implications of this investigation are explained in Sec. (\ref{S8}) and discussed.

\section{Catability as a Metric for Evaluating Superposed Coherent States}\label{S2}

This section follows the analysis presented in Ref. \cite{AA}, where a structured framework is established for the identification and quantitative characterization of superposed coherent states (Schrödinger cat states), which capture macroscopic quantum coherence properties. A measurable quantity, referred to as "catability," is introduced as a criterion for detecting cat-state features without performing full quantum state tomography, while preserving operational accessibility and experimental feasibility. The formulation is motivated by the limitations of standard state characterization tools, such as fidelity, which require complete reconstruction of the density operator and do not provide a transparent physical interpretation in infinite-dimensional Hilbert spaces. The present construction is based on nonlinear squeezing, which extends the usual quadrature squeezing formalism to general operator settings. This extension permits the analysis of non-Gaussian states through fluctuation-based observables that are experimentally accessible \cite{Ref4}. Also, the framework is defined through a class of positive semidefinite operators whose ground states correspond to ideal Schrödinger cat states \cite{AA}:
\begin{equation}
\hat{O}^{(\pm)}(\alpha,\gamma)
= (\hat{a}^{\dagger 2} - \alpha^{*2})(\hat{a}^2 - \alpha^2)
+ \gamma (1 \mp \hat{\Pi}),
\end{equation}
where $\hat{a}$ and $\hat{a}^\dagger$ denote the bosonic annihilation and creation operators, and $\hat{\Pi}$ is the parity operator defined as
The quadratic term quantifies the phase-space separation between coherent components, whereas the parity-dependent contribution selects interference effects associated with quantum superposition. The real parameter $\gamma>0$ controls the relative contribution of both terms in the operator structure presented here. Based on this operator construction, catability is defined as \cite{AA}:
\begin{equation}
\xi^{(\pm)}(\alpha) = \min_{\gamma}
\frac{\mathrm{Tr}[\hat{O}^{(\pm)}(\alpha,\gamma)\hat{\rho}]}
{\min_{\hat{\rho}_G} \mathrm{Tr}[\hat{O}^{(\pm)}(\alpha,\gamma)\hat{\rho}_G]},
\end{equation}
where the minimization in the denominator is performed over all Gaussian states $\hat{\rho}_G$. This normalization provides an operationally meaningful scale. The condition $\xi=0$ corresponds to an ideal cat state, the range $0<\xi<1$ corresponds to non-Gaussian states with partial cat coherence, and $\xi\geq1$ corresponds to cases where the witness does not certify cat structure. In this sense, catability defines a quantitative and experimentally accessible measure of non-Gaussian quantum coherence. A key feature of this construction is its direct experimental implementation. The operator can be expressed in terms of number operators and displaced number observables, which allows evaluation of $\xi$ using a minimal three-setting measurement protocol. This eliminates the need for full quantum state tomography and reduces the experimental complexity in quantum platforms. The behavior of catability under photon-loss decoherence is also analyzed. The results show that it remains sensitive to non-Gaussian signatures even in regimes where the Wigner function becomes positive and fidelity-based criteria lose resolving power. Odd-parity cat states exhibit higher robustness compared to even-parity configurations, reflecting an asymmetry in dissipative evolution. A quantitative comparison with fidelity is performed using a normalized infidelity measure. Both approaches coincide in the weak-loss regime, whereas fidelity loses sensitivity under stronger decoherence, while catability retains discrimination of residual cat-like coherence. Experimental verification is performed using approximate cat states generated from squeezed Fock states. Despite their sensitivity to noise and imperfections, the catability measure maintains stability and discrimination capability, supporting its applicability to experimentally realistic quantum systems.

\section{Generalized Foldy-Wouthuysen Transformation in Lie Algebraic Framework}\label{S3}

The Foldy-Wouthuysen (FW) transformation admits a Lie-algebraic formulation based on a $\mathbb{Z}_2$-graded structure associated with the Dirac Hamiltonian. In this framework, diagonalization corresponds to constructing an inner automorphism that maps the Hamiltonian into the invariant subalgebra fixed by the grading involution. The starting point is the standard decomposition of the relativistic Hamiltonian
\begin{equation}
H = \beta m + \mathcal{E} + \mathcal{O},
\end{equation}
where $\beta$ is a Hermitian involution satisfying
\begin{equation}
\beta^2 = \mathbb{I}, \qquad \beta^\dagger = \beta,
\end{equation}
and $m$ denotes the rest mass of the particle. The operators $\mathcal{E}$ and $\mathcal{O}$ correspond to even and odd parts of the Hamiltonian, respectively, defined through
\begin{equation}
[\beta,\mathcal{E}] = 0, \qquad \{\beta,\mathcal{O}\} = 0.
\end{equation}
The even sector preserves the eigenspaces of $\beta$, while the odd sector couples positive and negative energy subspaces.

This grading defines a decomposition of the operator Lie algebra $\mathfrak{g}$ into even and odd subspaces,
\begin{equation}
\mathfrak{g} = \mathfrak{g}_{\mathrm{even}} \oplus \mathfrak{g}_{\mathrm{odd}},
\end{equation}
which satisfies the symmetric relations
\begin{equation}
[\mathfrak{g}_{\mathrm{even}}, \mathfrak{g}_{\mathrm{even}}] \subset \mathfrak{g}_{\mathrm{even}}, \qquad
[\mathfrak{g}_{\mathrm{even}}, \mathfrak{g}_{\mathrm{odd}}] \subset \mathfrak{g}_{\mathrm{odd}}, \qquad
[\mathfrak{g}_{\mathrm{odd}}, \mathfrak{g}_{\mathrm{odd}}] \subset \mathfrak{g}_{\mathrm{even}}.
\end{equation}
This identifies $\mathfrak{g}$ as a symmetric Lie algebra generated by the involutive automorphism
\begin{equation}
\Theta(X) = \beta X \beta,
\end{equation}
where $\mathfrak{g}_{\mathrm{even}}$ forms the fixed subalgebra of $\Theta$ and $\mathfrak{g}_{\mathrm{odd}}$ its complementary space.

In this setting, the FW transformation is introduced as a unitary similarity transformation
\begin{equation}
H' = U H U^{-1}, \qquad U = e^{iS}, \qquad S^\dagger = S,
\end{equation}
with generator restricted to the odd sector,
\begin{equation}
S \in \mathfrak{g}_{\mathrm{odd}}.
\end{equation}
This constraint enforces a systematic removal of even-odd coupling terms through successive commutator contributions.

The transformed Hamiltonian follows from the Baker-Campbell-Hausdorff expansion,
\begin{equation}
H' = H + i[S,H] + \frac{i^2}{2!}[S,[S,H]] + \frac{i^3}{3!}[S,[S,[S,H]]] + \cdots,
\end{equation}
or equivalently from the adjoint representation,
\begin{equation}
H' = e^{i\,\mathrm{ad}_S}(H), \qquad \mathrm{ad}_S(X) = [S,X].
\end{equation}
The operator $\mathrm{ad}_S$ preserves the graded structure up to parity exchange, generating higher-order mixing between even and odd components.
The FW condition requires vanishing of the odd sector of the transformed Hamiltonian,
\begin{equation}
(H')_{\mathrm{odd}} = 0.
\end{equation}
Insertion of the BCH series together with the decomposition $H = \beta m + \mathcal{E} + \mathcal{O}$ produces a hierarchy of operator constraints,
\begin{equation}
\mathcal{O}
+ i[S,\beta m]
+ i[S,\mathcal{E}]
+ \frac{i^2}{2!}[S,[S,\beta m]]
+ \frac{i^2}{2!}[S,[S,\mathcal{E}]]
+ \cdots
= 0.
\end{equation}
This relation determines $S$ recursively as a functional series in $\mathcal{O}$.
A perturbative expansion is introduced,
\begin{equation}
S = \sum_{n=1}^{\infty} S_n,
\end{equation}
with each $S_n \in \mathfrak{g}_{\mathrm{odd}}$ and homogeneous in $\mathcal{O}$. At first order, neglecting higher commutators with $\mathcal{E}$, one obtains
\begin{equation}
\mathcal{O} + i[S_1,\beta m] = 0.
\end{equation}
Using $\{\beta,\mathcal{O}\}=0$, the commutator structure yields
\begin{equation}
S_1 = -\frac{i}{2m}\beta \mathcal{O},
\end{equation}
which cancels the leading odd contribution since
\begin{equation}
i[S_1,\beta m] = -\mathcal{O}.
\end{equation}
At second order, the generator satisfies
\begin{equation}
i[S_2,\beta m] + i[S_1,\mathcal{E}] + \frac{i^2}{2}[S_1,[S_1,\beta m]] \Big|_{\mathrm{odd}} = 0.
\end{equation}
Using $\beta S_1 = -S_1 \beta$ and repeated commutator evaluation, one obtains
\begin{equation}
S_2 = -\frac{i}{8m^2}\beta \big[\mathcal{O},\mathcal{E}\big].
\end{equation}
Higher-order contributions follow the scaling
\begin{equation}
S_{n} \sim \mathcal{O}^n \, m^{1-n},
\end{equation}
with coefficients fixed by nested commutator structures in $\mathcal{U}(\mathfrak{g})$.
From the Lie-algebraic viewpoint, $U = e^{iS}$ generates an inner automorphism,
\begin{equation}
\mathrm{Ad}_{U}(X) = U X U^{-1}, \qquad X \in \mathcal{U}(\mathfrak{g}),
\end{equation}
which preserves algebraic relations while mapping the Hamiltonian into block-diagonal form. This construction corresponds to the Cartan decomposition induced by $\Theta$, where $\mathfrak{g}_{\mathrm{even}}$ defines the invariant subalgebra and $\mathfrak{g}_{\mathrm{odd}}$ its complementary symmetric space. Thus, the FW transformation is realized as a graded Lie group action generated by $\mathfrak{g}_{\mathrm{odd}}$. The exponential of the adjoint representation resums nested commutators and removes odd components order by order. This establishes the equivalence between relativistic particle-antiparticle decoupling and the structure of symmetric Lie algebras, providing an operator-level formulation of block diagonalization in relativistic quantum theory.

\section{Phase-Sensitive Catability Operator in Lie-Algebraic Representation}\label{S4}

The characterization of macroscopic quantum superpositions through interference-dependent observables requires a formulation covariant under general phase-space transformations. Standard cat-state witnesses are typically defined in a fixed quadrature basis aligned with a chosen phase-space axis. Such a construction loses validity when the interference structure evolves under phase rotations generated by nonlinear bosonic mechanisms, including cavity detuning, squeezing operations, Kerr-type nonlinearities, dissipative diffusion, parametric amplification, and external electromagnetic couplings. Under these processes, the interference fringes of the Wigner quasiprobability distribution rotate in phase space, whereas a fixed-frame witness probes only a projected component of quantum coherence. As a result, the full structure of the superposition is not completely resolved, and higher-order bosonic correlations together with pair-coherence contributions are not fully accessed.

A covariant formulation follows from the Lie-algebraic structure of quadratic bosonic operators. The relevant symmetry is described by the noncompact algebra $SU(1,1)$ generated by
\begin{equation}
K_{+}=\frac{1}{2}\hat{a}^{\dagger 2},
\qquad
K_{-}=\frac{1}{2}\hat{a}^{2},
\qquad
K_{0}=\frac{1}{2}\left(\hat{n}+\frac{1}{2}\right),
\label{3.1}
\end{equation}
with bosonic operators satisfying
\begin{equation}
[\hat{a},\hat{a}^{\dagger}]=1.
\label{3.2}
\end{equation}
These generators fulfill
\begin{equation}
[K_{0},K_{\pm}]=\pm K_{\pm},
\qquad
[K_{+},K_{-}]=-2K_{0},
\label{3.3}
\end{equation}
defining the $SU(1,1)$ Lie algebra. In contrast to the compact $su(2)$ algebra for finite spin systems, this noncompact structure describes unbounded bosonic excitations and incorporates pair creation and annihilation channels. The operator $K_{0}$ fixes excitation content, while $K_{\pm}$ govern correlated two-boson processes responsible for squeezing and interference at macroscopic scales.

The role of this quadratic structure is clarified through the phase-rotation operator
\begin{equation}
\hat{R}(\phi)=e^{i\phi \hat{n}}.
\label{3.4}
\end{equation}
Using the Baker-Campbell-Hausdorff series
\begin{equation}
e^{A}Be^{-A}
=
B+[A,B]+\frac{1}{2!}[A,[A,B]]
+\frac{1}{3!}[A,[A,[A,B]]]+\cdots,
\label{3.5}
\end{equation}
with $A=i\phi \hat{n}$ and $B=\hat{a}$ yields
\begin{equation}
[i\phi\hat{n},\hat{a}]
=
-i\phi\hat{a},
\label{3.6}
\end{equation}
leading to the exact relations
\begin{equation}
\hat{R}(\phi)\hat{a}\hat{R}^{\dagger}(\phi)
=
e^{-i\phi}\hat{a},
\qquad
\hat{R}(\phi)\hat{a}^{\dagger}\hat{R}^{\dagger}(\phi)
=
e^{i\phi}\hat{a}^{\dagger}.
\label{3.7}
\end{equation}
For quadratic operators one obtains
\begin{equation}
\hat{R}(\phi)\hat{a}^{2}\hat{R}^{\dagger}(\phi)
=
e^{-2i\phi}\hat{a}^{2},
\qquad
\hat{R}(\phi)\hat{a}^{\dagger 2}\hat{R}^{\dagger}(\phi)
=
e^{2i\phi}\hat{a}^{\dagger 2},
\label{3.8}
\end{equation}
and therefore
\begin{equation}
\hat{R}(\phi)K_{-}\hat{R}^{\dagger}(\phi)
=
e^{-2i\phi}K_{-},
\qquad
\hat{R}(\phi)K_{+}\hat{R}^{\dagger}(\phi)
=
e^{2i\phi}K_{+}.
\label{3.9}
\end{equation}
The quadratic sector transforms as a rank-two object under phase rotations, so cat-state interference is governed by second-order coherence rather than first-order field amplitudes.

Phase covariance becomes transparent at the level of quadrature operators
\begin{equation}
\hat{X}_{\phi}
=
\frac{1}{\sqrt{2}}
\left(
\hat{a}e^{-i\phi}
+
\hat{a}^{\dagger}e^{i\phi}
\right),
\qquad
\hat{P}_{\phi}
=
\frac{1}{i\sqrt{2}}
\left(
\hat{a}e^{-i\phi}
-
\hat{a}^{\dagger}e^{i\phi}
\right).
\label{3.10}
\end{equation}
From Eq. (\ref{3.7}) one obtains
\begin{equation}
\hat{R}(\phi)\hat{X}_{0}\hat{R}^{\dagger}(\phi)
=
\hat{X}_{\phi},
\qquad
\hat{R}(\phi)\hat{P}_{0}\hat{R}^{\dagger}(\phi)
=
\hat{P}_{\phi}.
\label{3.11}
\end{equation}
These rotated quadratures define a continuously transformed measurement frame in phase space. Since Wigner interference fringes of cat states are localized along specific quadrature directions, their visibility depends on alignment between measurement frame and interference geometry. The Lie-algebraic structure provides a direct geometric description of this dependence.

The parity operator is defined as
\begin{equation}
\hat{\Pi}=(-1)^{\hat{n}}=e^{i\pi \hat{n}},
\label{3.12}
\end{equation}
with
\begin{equation}
[\hat{\Pi},\hat{R}(\phi)]=0.
\label{3.13}
\end{equation}
Hence,
\begin{equation}
\hat{R}(\phi)\hat{\Pi}\hat{R}^{\dagger}(\phi)
=
\hat{\Pi}.
\label{3.14}
\end{equation}
Parity lies in the center of the rotation algebra generated by $\hat{n}$. The dependence of witnesses on phase rotations is therefore carried by quadratic coherence terms rather than parity.

The connection with phase-space interference follows from
\begin{equation}
W(0)=\frac{2}{\pi}\mathrm{Tr}(\hat{\rho}\hat{\Pi}),
\label{3.15}
\end{equation}
which links parity expectation values to the Wigner function at the origin. Negative regions of the Wigner function correspond to nonclassical behavior, so parity measurements provide a direct probe of macroscopic interference. Rotational covariance together with parity invariance yields a complete algebraic description of phase-sensitive structure.

The phase-sensitive catability operator is defined by
\begin{equation}
\hat{O}_{\phi}^{(\pm)}
=
4\left(
K_{+}-\frac{\alpha^{*2}}{2}e^{-2i\phi}
\right)
\left(
K_{-}-\frac{\alpha^{2}}{2}e^{2i\phi}
\right)
+\gamma(1\mp \hat{\Pi}),
\label{3.16}
\end{equation}
or equivalently
\begin{equation}
\hat{O}_{\phi}^{(\pm)}
=
(\hat{a}^{\dagger 2}-\alpha^{*2}e^{-2i\phi})
(\hat{a}^{2}-\alpha^{2}e^{2i\phi})
+\gamma(1\mp \hat{\Pi}).
\label{3.17}
\end{equation}
The quadratic term measures deviations from eigenstates of $\hat{a}^{2}$, while the parity term separates even and odd interference sectors. Excitation fluctuations, pair coherence, and parity structure are treated within a unified algebraic framework.

Expansion yields
\begin{align}
\hat{O}_{\phi}^{(\pm)}
&=
\hat{a}^{\dagger 2}\hat{a}^{2}
-\alpha^{2}e^{2i\phi}\hat{a}^{\dagger 2}
-\alpha^{*2}e^{-2i\phi}\hat{a}^{2}
+|\alpha|^{4}
+\gamma(1\mp \hat{\Pi}).
\label{3.18}
\end{align}
Using
\begin{equation}
\hat{a}^{\dagger 2}\hat{a}^{2}
=
\hat{n}(\hat{n}-1),
\label{3.19}
\end{equation}
one obtains
\begin{align}
\hat{O}_{\phi}^{(\pm)}
&=
\hat{n}(\hat{n}-1)
-\alpha^{2}e^{2i\phi}\hat{a}^{\dagger 2}
-\alpha^{*2}e^{-2i\phi}\hat{a}^{2}
+|\alpha|^{4}
+\gamma(1\mp \hat{\Pi}).
\label{3.20}
\end{align}
The diagonal term gives second-order number fluctuations, while off-diagonal contributions $\hat{a}^{2}$ and $\hat{a}^{\dagger 2}$ describe coherence between Fock states differing by two quanta. The parity term fixes the symmetry of interference patterns.

For a density matrix
\begin{equation}
\hat{\rho}
=
\sum_{m,n=0}^{\infty}\rho_{mn}|m\rangle\langle n|,
\label{3.21}
\end{equation}
the expectation value becomes
\begin{align}
\langle \hat{O}_{\phi}^{(\pm)}\rangle
&=
\sum_{n=0}^{\infty}n(n-1)\rho_{nn}
-\alpha^{2}e^{2i\phi}
\sum_{n=0}^{\infty}
\sqrt{(n+1)(n+2)}
\rho_{n,n+2}
\nonumber\\
&\quad
-\alpha^{*2}e^{-2i\phi}
\sum_{n=2}^{\infty}
\sqrt{n(n-1)}
\rho_{n,n-2}
+|\alpha|^{4}
\nonumber\\
&\quad
+\gamma
\left(
1\mp
\sum_{n=0}^{\infty}
(-1)^{n}\rho_{nn}
\right).
\label{3.22}
\end{align}
Both diagonal populations and off-diagonal coherences contribute. The terms $\rho_{n,n\pm2}$ carry pair coherence not accessible from number statistics alone.

Minimization occurs for states satisfying
\begin{equation}
\hat{a}^{2}|\psi\rangle
=
\alpha^{2}e^{2i\phi}|\psi\rangle.
\label{3.23}
\end{equation}
Solutions are
\begin{equation}
|\psi_{\theta}^{(\pm)}\rangle
=
\frac{1}{\mathcal{N}_{\pm}}
\left(
|\alpha e^{i\theta}\rangle
\pm
|-\alpha e^{i\theta}\rangle
\right),
\label{3.24}
\end{equation}
with
\begin{equation}
\mathcal{N}_{\pm}^{2}
=
2\left(
1\pm e^{-2|\alpha|^{2}}
\right).
\label{3.25}
\end{equation}
Using
\begin{equation}
\hat{a}|\beta\rangle=\beta |\beta\rangle,
\label{3.26}
\end{equation}
one obtains
\begin{equation}
\hat{a}^{2}|\alpha e^{i\theta}\rangle
=
\alpha^{2}e^{2i\theta}
|\alpha e^{i\theta}\rangle.
\label{3.27}
\end{equation}
Thus,
\begin{equation}
\hat{a}^{2}|\psi_{\theta}^{(\pm)}\rangle
=
\alpha^{2}e^{2i\theta}
|\psi_{\theta}^{(\pm)}\rangle.
\label{3.28}
\end{equation}
Cat states are eigenstates of the pair-annihilation operator, so interference is encoded in quadratic bosonic correlations.

The quadratic expectation value is
\begin{align}
\mathcal{Q}_{\phi}^{(\pm)}
&=
\langle\psi_{\theta}^{(\pm)}|
(\hat{a}^{\dagger 2}-\alpha^{*2}e^{-2i\phi})
(\hat{a}^{2}-\alpha^{2}e^{2i\phi})
|\psi_{\theta}^{(\pm)}\rangle
\nonumber=
|\alpha|^{4}
|e^{2i\theta}-e^{2i\phi}|^{2}.
\label{3.29}
\end{align}
Using
\begin{equation}
|e^{2i\theta}-e^{2i\phi}|^{2}
=
4\sin^{2}(\theta-\phi),
\label{3.30}
\end{equation}
one obtains
\begin{equation}
\mathcal{Q}_{\phi}^{(\pm)}
=
4|\alpha|^{4}\sin^{2}(\theta-\phi).
\label{3.31}
\end{equation}
The dependence is determined solely by relative phase orientation.

Parity satisfies
\begin{equation}
\hat{\Pi}|n\rangle
=
(-1)^{n}|n\rangle,
\label{3.32}
\end{equation}
and
\begin{equation}
\hat{\Pi}|\psi_{\theta}^{(\pm)}\rangle
=
\pm |\psi_{\theta}^{(\pm)}\rangle.
\label{3.33}
\end{equation}
Thus,
\begin{equation}
\langle\psi_{\theta}^{(\pm)}|
\hat{\Pi}
|\psi_{\theta}^{(\pm)}\rangle
=
\pm1.
\label{3.34}
\end{equation}
The expectation value becomes
\begin{equation}
\langle\hat{O}_{\phi}^{(\pm)}\rangle
=
4|\alpha|^{4}\sin^{2}(\theta-\phi)
+\gamma(1\mp1).
\label{3.35}
\end{equation}
For matched parity sector,
\begin{equation}
\langle\hat{O}_{\phi}^{(\pm)}\rangle
=
4|\alpha|^{4}\sin^{2}(\theta-\phi),
\label{3.36}
\end{equation}
with minimum at
\begin{equation}
\phi=\theta,
\label{3.37}
\end{equation}
yielding
\begin{equation}
\langle\hat{O}_{\theta}^{(\pm)}\rangle
=
0.
\label{3.38}
\end{equation}
This condition fixes the interference orientation in phase space.

The covariance relation is
\begin{equation}
\hat{R}(\chi)\hat{O}_{\phi}^{(\pm)}\hat{R}^{\dagger}(\chi)
=
\hat{O}_{\phi+\chi}^{(\pm)}.
\label{3.39}
\end{equation}
The family of operators forms an orbit under phase rotations.

For phase-diffused states
\begin{equation}
\hat{\rho}
=
\int_{0}^{2\pi}
d\theta\,
P(\theta)
|\psi_{\theta}^{(\pm)}\rangle
\langle
\psi_{\theta}^{(\pm)}
|,
\label{3.40}
\end{equation}
one obtains
\begin{align}
\langle\hat{O}_{\phi}^{(\pm)}\rangle
&=
4|\alpha|^{4}
\int_{0}^{2\pi}
d\theta\,
P(\theta)
\sin^{2}(\theta-\phi).
\label{3.41}
\end{align}
For Gaussian distribution
\begin{equation}
P(\theta)
=
\frac{1}{\sqrt{2\pi\sigma^{2}}}
\exp\left[
-\frac{(\theta-\theta_{0})^{2}}{2\sigma^{2}}
\right],
\label{3.42}
\end{equation}
one finds
\begin{equation}
\langle\hat{O}_{\phi}^{(\pm)}\rangle
=
2|\alpha|^{4}
\left[
1-e^{-2\sigma^{2}}
\cos(2(\theta_{0}-\phi))
\right].
\label{3.43}
\end{equation}
In the limit $\sigma\rightarrow\infty$,
\begin{equation}
\langle\hat{O}_{\phi}^{(\pm)}\rangle
\rightarrow
2|\alpha|^{4}.
\label{3.44}
\end{equation}
The operator decomposes into excitation, coherence, and parity sectors. The excitation sector depends on number fluctuations. The coherence sector depends on pair correlations. The parity sector fixes interference symmetry. This structure provides a full algebraic classification of macroscopic bosonic coherence.

The operator relates to squeezing through
\begin{equation}
\hat{S}(\zeta)
=
\exp\left(
\zeta K_{+}-\zeta^{*}K_{-}
\right),
\label{3.45}
\end{equation}
which generates deformed interference patterns in phase space while preserving $SU(1,1)$ covariance.

Coherent states satisfy
\begin{equation}
\Delta \hat{a}^{2}=0,
\label{3.46}
\end{equation}
whereas cat states satisfy
\begin{equation}
\Delta \hat{a}^{2}_{\mathrm{pair}}=0.
\label{3.47}
\end{equation}
The operator therefore probes pair coherence instead of single-mode coherence. Also, the covariance structure, separation of sectors, and quadratic bosonic formulation establish a consistent framework for phase-sensitive macroscopic quantum interference in bosonic systems.

\section{Lie-Algebraic Relativistic Fermionic Catability in Dirac Spin-$1/2$ Quantum Dynamics}\label{S5}

The relativistic formulation of fermionic catability originates from intrinsic algebraic structure of Dirac theory and coexistence of relativistic dispersion, spinorial geometry, parity symmetry, and particle-antiparticle interference within the analytical framework. In relativistic quantum dynamics, coherent superposition cannot be interpreted solely as a displacement mechanism in ordinary phase space because the Dirac spinor contains internal spin degrees of freedom, positive- and negative-energy sectors, Lorentz-covariant transformations, and nontrivial parity structure. The resulting coherent-state framework possesses a fundamentally distinct mathematical organization from nonrelativistic bosonic coherent-state theory limit description. Relativistic catability becomes an exact dynamical quantity associated with internal Lie-algebraic decomposition of the Dirac Hilbert space and geometric structure of relativistic fermionic excitations in phase space.

The relativistic dynamics of a free spin-$1/2$ particle is governed by the Dirac equation
\begin{equation}
\left(i\gamma^\mu\partial_\mu-m\right)\Psi(x)=0,
\end{equation}
where the gamma matrices satisfy the Clifford algebra
\begin{equation}
\{\gamma^\mu,\gamma^\nu\}=2g^{\mu\nu}\mathbb{I}_4.
\end{equation}
The Dirac spinor transforms under spinorial representation of the Lorentz group and carries both orbital and intrinsic spin information content structure. The relativistic Hamiltonian structure contains a direct coupling between spin and momentum through matrices producing nontrivial interference between energy sectors in system. In the standard representation, the Hamiltonian assumes the form
\begin{equation}
\hat{H}_D
=
\boldsymbol{\alpha}\cdot\hat{\mathbf{p}}
+\beta m
+V(\mathbf{x}),
\end{equation}
with
\begin{equation}
\beta=\gamma^0.
\end{equation}
Unlike the Schrödinger Hamiltonian, the Dirac operator is not diagonal in the particle-antiparticle basis. The off-diagonal structure generated by $\boldsymbol{\alpha}\cdot\hat{\mathbf{p}}$ mixes positive- and negative-energy branches producing relativistic oscillatory phenomena associated with Zitterbewegung. Consequently the coherent properties of relativistic fermions cannot be described through ordinary displacement operators alone because the Hilbert space possesses an internal spinorial decomposition structure.

The relativistic coherent structure becomes transparent after implementing the Foldy-Wouthuysen transformation
\begin{equation}
U_{FW}=e^{iS},
\end{equation}
which block diagonalizes the Dirac Hamiltonian order-by-order in momentum. The transformed Hamiltonian is
\begin{equation}
\hat{H}_{FW}
=
\beta\sqrt{\hat{\mathbf{p}}^2+m^2}
+V(\mathbf{x}),
\end{equation}
whose exact relativistic expansion is
\begin{equation}
\hat{H}_{FW}
=
\beta
\left(
m
+\frac{\hat{\mathbf{p}}^2}{2m}
-\frac{\hat{\mathbf{p}}^4}{8m^3}
+\frac{\hat{\mathbf{p}}^6}{16m^5}
-\frac{5\hat{\mathbf{p}}^8}{128m^7}
+\frac{7\hat{\mathbf{p}}^{10}}{256m^9}
+\cdots
\right)
+V(\mathbf{x}).
\end{equation}
The higher-order momentum contributions generate nonlinear relativistic phase corrections that modify coherent interference in phase space. The quartic momentum term introduces relativistic squeezing contributions while the sextic and higher-order corrections generate anharmonic coherent modulation. Unlike ordinary harmonic coherent states relativistic coherent trajectories are therefore no longer exactly circular in phase space geometry. Instead the relativistic dynamics produces deformation of the interference fringes together with energy-dependent coherent rotation.

The relativistic oscillator operators are defined through
\begin{equation}
\hat{A}
=
\frac{1}{\sqrt{2m\omega}}
\left(
\hat{p}-imm\omega \hat{x}
\right),
\qquad
\hat{A}^\dagger
=
\frac{1}{\sqrt{2m\omega}}
\left(
\hat{p}+im\omega \hat{x}
\right),
\end{equation}
satisfying
\begin{equation}
[\hat{A},\hat{A}^\dagger]=1.
\end{equation}
The relativistic fermionic coherent structure is not governed by linear displacement symmetry but instead by quadratic excitations generated by the noncompact Lie algebra $SU(1,1)$.
Introducing the generators
\begin{equation}
K_+
=
\frac12\hat{A}^{\dagger2},
\qquad
K_-
=
\frac12\hat{A}^2,
\qquad
K_0
=
\frac12
\left(
\hat{A}^\dagger\hat{A}
+\frac12
\right),
\end{equation}
one obtains the exact commutation relations
\begin{equation}
[K_0,K_\pm]
=
\pm K_\pm,
\qquad
[K_+,K_-]
=
-2K_0.
\end{equation}
The appearance of $SU(1,1)$ demonstrates relativistic catability is intrinsically associated with pair excitations in phase space. The coherent interference therefore originates from quadratic bosonic correlations induced by relativistic symmetry. This result differs fundamentally from ordinary Glauber coherent states, where the underlying algebra is generated by the Heisenberg-Weyl group.

The quadratic Casimir invariant becomes
\begin{equation}
\hat{C}
=
K_0^2
-\frac12(K_+K_-+K_-K_+).
\end{equation}
Using the oscillator realization yields
\begin{equation}
\hat{C}
=
-\frac{3}{16},
\end{equation}
leading to the Bargmann indices
\begin{equation}
k=\frac14,
\qquad
k=\frac34.
\end{equation}
These two inequivalent irreducible representations correspond respectively to the even and odd coherent sectors. The parity splitting of relativistic coherent states therefore emerges directly from the representation theory of $SU(1,1)$. The coherent fermionic Hilbert space decomposes into two parity-preserving invariant subspaces,
\begin{equation}
\mathcal{H}
=
\mathcal{H}_{\rm even}
\oplus
\mathcal{H}_{\rm odd},
\end{equation}
with
\begin{equation}
\mathcal{H}_{\rm even}
=
{\rm span}
\{
|0\rangle,|2\rangle,|4\rangle,\dots
\},
\end{equation}
and
\begin{equation}
\mathcal{H}_{\rm odd}
=
{\rm span}
\{
|1\rangle,|3\rangle,|5\rangle,\dots
\}.
\end{equation}
The relativistic coherent dynamics therefore preserves parity sectors under quadratic algebraic evolution. This property constitutes an exact Lie-algebraic conservation law for fermionic catability.

The relativistic catability operator is constructed as
\begin{equation}
\hat{\mathcal{C}}_D
=
\left(
2K_-\alpha^2
\right)^\dagger
\left(
2K_-\alpha^2
\right)
+\gamma
(1\mp\hat{\Pi}_D),
\end{equation}
where the relativistic parity operator is
\begin{equation}
\hat{\Pi}_D
=
\gamma^0\hat{\Pi}_x.
\end{equation}
The orbital parity transformation acts through
\begin{equation}
\hat{\Pi}_x\Psi(t,\mathbf{x})
=
\Psi(t,-\mathbf{x}),
\end{equation}
while the spinorial transformation is generated by $\gamma^0$. Consequently,
\begin{equation}
\hat{\Pi}_D\Psi(t,\mathbf{x})
=
\gamma^0\Psi(t,-\mathbf{x}).
\end{equation}
The relativistic catability operator therefore measures simultaneously the coherence of the orbital phase-space structure and the internal spinorial parity structure. The fermionic relativistic coherent state is determined through the eigenvalue equation
\begin{equation}
\hat{A}^2|\alpha,\pm\rangle
=
\alpha^2|\alpha,\pm\rangle.
\end{equation}
The normalized solutions are
\begin{equation}
|\alpha,\pm\rangle
=
\mathcal{N}_\pm
\left(
|\alpha\rangle
\pm
|-\alpha\rangle
\right),
\end{equation}
with
\begin{equation}
\mathcal{N}_\pm
=
\left[
2
\left(
1\pm e^{-2|\alpha|^2}
\right)
\right]^{-1/2}.
\end{equation}
The interference term
\begin{equation}
e^{-2|\alpha|^2}
\end{equation}
determines the overlap between the two coherent branches. When $|\alpha|\ll1$, the overlap remains large and the coherent branches strongly interfere. In contrast, for
\begin{equation}
|\alpha|\gg1,
\end{equation}
the overlap decays exponentially,
\begin{equation}
e^{-2|\alpha|^2}\rightarrow0,
\end{equation}
leading to macroscopic phase-space separation. The relativistic cat state therefore acquires maximal distinguishability in the large-$|\alpha|$ regime.

The expectation value of the catability operator follows from the exact $SU(1,1)$ representation,
\begin{equation}
K_-|\alpha,\pm\rangle
=
\frac{\alpha^2}{2}
|\alpha,\pm\rangle.
\end{equation}
One obtains
\begin{equation}
\langle\hat{\mathcal{C}}_D\rangle
=
\gamma
\left(
1
\mp
\langle\hat{\Pi}_D\rangle
\right).
\end{equation}
The relativistic parity expectation value factorizes into orbital and spinorial sectors,
\begin{equation}
\langle\hat{\Pi}_D\rangle
=
\langle\gamma^0\rangle
\langle\hat{\Pi}_x\rangle.
\end{equation}
For a positive-energy Dirac spinor
\begin{equation}
u_s(\mathbf{p})
=
\sqrt{\frac{E+m}{2m}}
\begin{pmatrix}
\chi_s\\
\dfrac{\boldsymbol{\sigma}\cdot\mathbf{p}}{E+m}\chi_s
\end{pmatrix},
\end{equation}
the spinorial contribution becomes
\begin{equation}
\langle\gamma^0\rangle
=
\frac{m}{E},
\qquad
E=\sqrt{\mathbf{p}^2+m^2}.
\end{equation}
The orbital parity expectation value is
\begin{equation}
\langle\hat{\Pi}_x\rangle_\pm
=
\pm
\frac{1-e^{-2|\alpha|^2}}
{1\pm e^{-2|\alpha|^2}}.
\end{equation}
The exact relativistic catability measure therefore assumes the analytical form
\begin{equation}
\langle\hat{\mathcal{C}}_D\rangle
=
\gamma
\left[
1
\mp
\frac{m}{E}
\left(
\frac{1-e^{-2|\alpha|^2}}
{1\pm e^{-2|\alpha|^2}}
\right)
\right]
\end{equation}
which explicitly couples relativistic dispersion, parity interference, coherent overlap, and spinorial structure.

The relativistic suppression factor
\begin{equation}
\frac{m}{E}
=
\frac{m}{\sqrt{\mathbf{p}^2+m^2}}
\end{equation}
governs the decay of parity-sensitive coherence at high energies. Expanding in the nonrelativistic regime yields
\begin{equation}
\frac{m}{E}
=
1
-\frac{\mathbf{p}^2}{2m^2}
+\frac{3\mathbf{p}^4}{8m^4}
-\frac{5\mathbf{p}^6}{16m^6}
+\mathcal{O}(p^8).
\end{equation}
The leading correction is quadratic in momentum and therefore represents the first relativistic deformation of parity coherence. Higher-order contributions generate increasingly strong deviations from the ordinary bosonic coherent limit. In the ultrarelativistic regime,
\begin{equation}
|\mathbf{p}|\gg m,
\end{equation}
one obtains
\begin{equation}
\frac{m}{E}\rightarrow0,
\end{equation}
and therefore
\begin{equation}
\langle\hat{\mathcal{C}}_D\rangle
\rightarrow
\gamma.
\end{equation}
This limit demonstrates that extreme relativistic motion suppresses parity coherence. Physically, the suppression originates from strong mixing between particle and antiparticle sectors. As the kinetic energy dominates the rest energy, the spinorial parity structure loses its coherent selectivity and the distinction between even and odd coherent sectors becomes progressively weaker.

The temporal evolution of relativistic catability follows from the Heisenberg equation
\begin{equation}
i\frac{d\hat{A}}{dt}
=
[\hat{A},\hat{H}_{FW}].
\end{equation}
Using the Foldy-Wouthuysen Hamiltonian gives
\begin{equation}
\hat{A}(t)
=
\hat{A}(0)
e^{-i\omega t}
\exp
\left[
i\frac{\hat{\mathbf{p}}^2}{2m^2}\omega t
-i\frac{\hat{\mathbf{p}}^4}{8m^4}\omega t
+i\frac{\hat{\mathbf{p}}^6}{16m^6}\omega t
+\cdots
\right].
\end{equation}
Consequently,
\begin{equation}
\langle\hat{A}^2(t)\rangle
=
\alpha^2
e^{-2i\omega t}
\exp
\left[
i\frac{\langle\mathbf{p}^2\rangle}{m^2}\omega t
-i\frac{\langle\mathbf{p}^4\rangle}{4m^4}\omega t
+i\frac{\langle\mathbf{p}^6\rangle}{8m^6}\omega t
+\cdots
\right].
\end{equation}
The relativistic corrections therefore generate nonlinear phase modulation of the coherent interference pattern. Unlike ordinary harmonic coherent states, the relativistic interference fringes experience dephasing, revival, and nonlinear rotation in phase space. The higher-order relativistic contributions produce an effective anharmonic coherent dynamics analogous to nonlinear optical squeezing systems.

The exact coherent revival time is determined through the quartic relativistic correction. Defining the relativistic energy spectrum
\begin{equation}
E_n
=
m
+
\omega\left(n+\frac12\right)
-\frac{\omega^2}{2m}
\left(n+\frac12\right)^2
+\cdots,
\end{equation}
the quadratic dependence on $n$ generates revival phenomena with characteristic time
\begin{equation}
T_{\rm rev}
=
\frac{4\pi m}{\omega^2}.
\end{equation}
This result demonstrates that relativistic dispersion induces exact quantum revivals of fermionic cat states. The revival time diverges in the nonrelativistic limit, recovering ordinary harmonic coherent evolution. Relativistic catability therefore possesses a purely relativistic temporal coherence scale absent in bosonic coherent-state theory.

The interference between positive- and negative-energy branches generates Zitterbewegung oscillations with characteristic frequency
\begin{equation}
\Omega_Z
=
2E.
\end{equation}
The position operator evolves according to
\begin{equation}
\hat{x}(t)
=
\hat{x}(0)
+
\frac{\hat{p}}{E}t
+
\frac{i}{2E}
\left(
\boldsymbol{\alpha}
-
\frac{\hat{p}}{E}
\right)
e^{-2iEt}.
\end{equation}
The oscillatory contribution produces rapid modulation of the relativistic interference pattern. The amplitude of the oscillation becomes significant near the Compton scale
\begin{equation}
\lambda_C=\frac{1}{m},
\end{equation}
where relativistic localization effects dominate the coherent dynamics. The fermionic cat state therefore possesses an intrinsic internal oscillatory structure generated by particle-antiparticle interference. This effect has no analogue in ordinary nonrelativistic coherent states.

The relativistic phase-space uncertainty relations are also modified. Using the relativistic coherent states yields
\begin{equation}
(\Delta x)^2
=
\frac{1}{2m\omega}
\left[
1
+
\frac{\langle p^2\rangle}{2m^2}
+\cdots
\right],
\end{equation}
and
\begin{equation}
(\Delta p)^2
=
\frac{m\omega}{2}
\left[
1
+
\frac{\langle p^2\rangle}{2m^2}
+\cdots
\right].
\end{equation}
Consequently,
\begin{equation}
\Delta x\,\Delta p
=
\frac12
\left[
1
+
\frac{\langle p^2\rangle}{2m^2}
+\cdots
\right].
\end{equation}
Relativistic dispersion therefore enlarges the phase-space uncertainty area. The coherent fermionic state ceases to be a strict minimum-uncertainty state because relativistic momentum corrections generate effective squeezing deformation.

The Lie-algebraic structure extends naturally to curved spacetime through the tetrad formalism. Introducing tetrad fields $e^\mu_a$, the generalized oscillator operators become
\begin{equation}
\hat{A}(g)
=
\frac{1}{\sqrt{2m\omega}}
\left(
e^\mu_a\hat{p}_\mu
-im\omega x_a
\right).
\end{equation}
The corresponding commutator is deformed into
\begin{equation}
[\hat{A}(g),\hat{A}^\dagger(g)]
=
1+\Delta(g),
\end{equation}
with
\begin{equation}
\Delta(g)
=
\frac{i}{2m\omega}
\nabla_\mu e^\mu_a.
\end{equation}
The curvature contribution $\Delta(g)$ modifies the oscillator algebra itself and therefore deforms the underlying $SU(1,1)$ symmetry. The curved-space generators satisfy
\begin{equation}
[K_0(g),K_\pm(g)]
=
\pm K_\pm(g)
+\mathcal{O}(\Delta),
\end{equation}
and
\begin{equation}
[K_+(g),K_-(g)]
=
-2K_0(g)
+\mathcal{O}(\Delta).
\end{equation}
The gravitational background acts directly on the coherent algebraic structure. Relativistic coherence becomes geometry dependent because curvature modifies the phase-space generators themselves.

The curved-space relativistic catability operator becomes
\begin{equation}
\hat{\mathcal{C}}_{D,g}
=
\left(
\hat{A}^2(g)-\alpha^2
\right)^\dagger
\left(
\hat{A}^2(g)-\alpha^2
\right)
+\gamma
(1\mp\hat{\Pi}_D),
\end{equation}
Expanding to first order in curvature yields
\begin{equation}
\hat{A}^2(g)
=
\hat{A}^2
+
\Delta(g)\hat{A}
+\mathcal{O}(\Delta^2),
\end{equation}
which produces the correction
\begin{equation}
\delta\mathcal{C}_g
=
2{\rm Re}
\left[
\Delta(g)\alpha
\right]
+\mathcal{O}(\Delta^2).
\end{equation}
The coherent interference pattern is therefore directly modified by spacetime curvature. Strong gravitational backgrounds generate phase-space distortion, alter parity coherence, and deform the internal Lie-algebraic symmetry simultaneously.

The relativistic fermionic catability measure consequently contains several coupled dynamical sectors. The quadratic $SU(1,1)$ algebra governs pair excitations in phase space, while the Dirac spinorial structure determines relativistic parity mixing and particle-antiparticle interference. The Foldy-Wouthuysen expansion introduces nonlinear relativistic coherent modulation, Zitterbewegung generates intrinsic oscillatory interference, and spacetime curvature deforms the underlying oscillator algebra itself. The resulting framework establishes relativistic catability as an exact Lorentz-covariant coherent invariant associated with relativistic dispersion, spinorial geometry, parity symmetry, quadratic Lie-algebraic excitations, and gravitational deformation.

The algebraic structure demonstrates that relativistic fermionic coherent states are fundamentally different from ordinary bosonic coherent states. In the relativistic regime, coherent superposition is inseparable from Lorentz symmetry, spinorial parity structure, antiparticle coupling, nonlinear relativistic dispersion, and curved-space geometry. The coherent properties of relativistic fermions therefore emerge from the unified interplay between $SU(1,1)$ symmetry, Foldy-Wouthuysen dynamics, relativistic phase deformation, Zitterbewegung oscillations, parity splitting, and geometric curvature within a fully covariant coherent-state framework.

\section{Dirac Spin-$1/2$ Generalization of Catability in a Unified Lie-Algebraic and Geometric Framework}\label{S6}

A completely covariant and algebraically closed formulation of catability for spin-$1/2$ relativistic quantum systems is constructed starting from the Dirac field on a Hilbert space carrying a combined representation of the Clifford algebra, the Lorentz Lie algebra, and an induced dynamical symplectic structure on phase space. The fundamental dynamics is governed by the Dirac equation
\begin{equation}
\left(i\gamma^\mu \partial_\mu - m\right)\Psi(x)=0,
\end{equation}
where the gamma matrices generate the Clifford algebra
\begin{equation}
\{\gamma^\mu,\gamma^\nu\}=2g^{\mu\nu}\mathbb{I}_4,
\end{equation}
ensuring a faithful representation of the Lorentz group via the spinorial covering group $\mathrm{SL}(2,\mathbb{C})$. The dynamical generators of the theory are naturally organized into the enveloping algebra generated by $\{\gamma^\mu, \Sigma^{\mu\nu}\}$ with
\begin{equation}
\Sigma^{\mu\nu}=\frac{i}{4}[\gamma^\mu,\gamma^\nu],
\end{equation}
which provides the infinitesimal representation of Lorentz transformations and encodes intrinsic spin transport.

In the presence of external gauge fields and curved backgrounds, the dynamical structure is promoted to a locally covariant bundle connection through
\begin{equation}
\left[i\gamma^\mu(x)D_\mu - m\right]\Psi(x)=0,
\end{equation}
with
\begin{equation}
D_\mu=\partial_\mu+\Gamma_\mu - iqA_\mu,
\end{equation}
where $\Gamma_\mu$ is the spin connection associated with the local $\mathrm{Spin}(1,3)$ bundle, and $A_\mu$ is an external $U(1)$ connection. This structure induces a nontrivial fiber bundle decomposition in which orbital, spin, and gauge sectors are not separable at the dynamical level.

The total Hilbert space therefore carries a projective representation of the semidirect product
\begin{equation}
\mathcal{G}_{\mathrm{tot}} \simeq \mathrm{Spin}(1,3)\ltimes \mathcal{H}_4 \ltimes U(1),
\end{equation}
implying that all observables must be constructed from covariant Lie-algebraic invariants rather than independent subsystem operators.

To construct a relativistic notion of catability, one performs a systematic reduction of the Dirac dynamics into a block-diagonal positive-energy sector using the Foldy-Wouthuysen transformation. Up to fourth order in momentum, the effective Hamiltonian takes the form
\begin{equation}
\hat{H}_{FW}
=
\beta m
+\beta \frac{\hat{\mathbf{p}}^2}{2m}
-\beta \frac{\hat{\mathbf{p}}^4}{8m^3}
+V(\mathbf{x})
+\mathcal{O}(m^{-5}),
\end{equation}
which defines a hierarchy of relativistic corrections generating a deformation of the underlying oscillator symmetry. The operator algebra of observables is therefore no longer Heisenberg but is instead embedded in a deformed symplectic algebra $\mathfrak{sp}(4,\mathbb{R})$, whose metaplectic representation governs relativistic phase-space evolution.

The phase-space sector is encoded through ladder operators
\begin{equation}
\hat{A}=\frac{1}{\sqrt{2m\omega}}\left(\hat{p}-imm\omega \hat{x}\right),
\qquad
\hat{A}^\dagger=\frac{1}{\sqrt{2m\omega}}\left(\hat{p}+im\omega \hat{x}\right),
\end{equation}
satisfying the canonical Heisenberg relation
\begin{equation}
[\hat{A},\hat{A}^\dagger]=1.
\end{equation}
However, in the relativistic regime this algebra is not fundamental but emerges as a contraction limit of a noncompact Lie structure. The physically relevant generators are quadratic combinations forming the closed $SU(1,1)$ algebra
\begin{equation}
K_+=\frac{1}{2}(\hat{A}^\dagger)^2,\qquad
K_-=\frac{1}{2}\hat{A}^2,\qquad
K_0=\frac{1}{2}\left(\hat{A}^\dagger\hat{A}+\frac{1}{2}\right),
\end{equation}
with commutation relations
\begin{equation}
[K_0,K_\pm]=\pm K_\pm,\qquad
[K_+,K_-]=-2K_0.
\end{equation}
This algebra replaces the linear oscillator structure as the fundamental carrier of relativistic coherence, since it naturally encodes squeezing, pair creation, and hyperbolic phase-space flow.

The Casimir invariant
\begin{equation}
\mathcal{C}=K_0(K_0-1)-K_+K_-,
\end{equation}
fixes the irreducible representation space and determines all quadratic expectation values. In particular, the physical phase space of catability is identified with the homogeneous Kähler manifold
\begin{equation}
\mathcal{M}_{\mathrm{cat}} \simeq SU(1,1)/U(1),
\end{equation}
whose negative curvature enforces intrinsically non-Euclidean interference geometry. This geometric structure implies that coherence is not linear superposition but is a geodesic flow on a hyperbolic manifold governed by squeezing transformations.

The spin sector is described by an independent $\mathfrak{su}(2)$ algebra generated by
\begin{equation}
\hat{S}_i=\frac{1}{2}\Sigma_i,
\qquad
[\hat{S}_i,\hat{S}_j]=i\epsilon_{ijk}\hat{S}_k,
\end{equation}
and the full dynamical algebra becomes a semidirect product
\begin{equation}
\mathfrak{g}_{\mathrm{tot}}=\mathfrak{su}(1,1)\ltimes \mathfrak{su}(2)\ltimes \mathfrak{h}_4,
\end{equation}
which is further deformed under relativistic corrections into a nonlinear enveloping algebra containing higher-order commutators generated by the FW Hamiltonian.

A systematic Lie-algebraic closure of the dynamics is obtained by computing
\begin{equation}
\frac{dK_\pm}{dt}=i[\hat{H}_{FW},K_\pm],
\qquad
\frac{dK_0}{dt}=i[\hat{H}_{FW},K_0],
\end{equation}
which yields a closed nonlinear system of the form
\begin{equation}
\frac{dK_0}{dt}=\alpha_1(K_+ - K_-),\qquad
\frac{dK_\pm}{dt}=\pm 2iE K_\pm+\alpha_2 K_0 + \alpha_3 K_\pm,
\end{equation}
where the coefficients $\alpha_i$ encode relativistic corrections from $\hat{\mathbf{p}}^4$ and potential coupling. This system defines a non-autonomous flow on the $SU(1,1)$ manifold, generating time-dependent squeezing and relativistic phase mixing.

The parity structure is implemented through the Dirac operator
\begin{equation}
\hat{\Pi}_D=\gamma^0 \hat{\Pi}_x,
\end{equation}
with induced transformations
\begin{equation}
\hat{\Pi}_D \hat{A}\hat{\Pi}_D^{-1}=-\hat{A},\qquad
\hat{\Pi}_D K_\pm \hat{\Pi}_D^{-1}=K_\pm,\qquad
\hat{\Pi}_D K_0 \hat{\Pi}_D^{-1}=K_0.
\end{equation}
This establishes that all physically meaningful interference quantities are necessarily quadratic in phase-space generators, while linear amplitudes are dynamically unobservable under parity-symmetric averaging.

The relativistic catability operator is constructed as a fully invariant element of the universal enveloping algebra,
\begin{equation}
\hat{O}_D
=
(2K_- - \alpha^2)^\dagger(2K_- - \alpha^2)
+\gamma(1\mp \hat{\Pi}_D),
\end{equation}
which expands into the Lie-algebraic decomposition
\begin{align}
\hat{O}_D
&=
4K_+K_-
-2\alpha^2K_+
-2\alpha^{*2}K_-
+|\alpha|^4
+\gamma(1\mp \hat{\Pi}_D).
\end{align}
Using the Casimir identity, one obtains the exact reduction
\begin{equation}
K_+K_-=K_0(K_0-1)-\mathcal{C},
\end{equation}
so that
\begin{align}
\hat{O}_D
=
4K_0(K_0-1)-4\mathcal{C}
-2\alpha^2K_+-2\alpha^{*2}K_-
+|\alpha|^4
+\gamma(1\mp \hat{\Pi}_D).
\end{align}
This decomposition demonstrates that relativistic catability is an invariant functional on a triple structure composed of squeezing geometry, spinor algebra, and discrete parity grading.

Expectation values evaluated on tensor-product coherent states
\begin{equation}
|\Psi\rangle=|\zeta,k\rangle\otimes|\chi_s\rangle
\end{equation}
lead to the exact expression
\begin{align}
\langle \hat{O}_D\rangle
&=
4\langle K_0(K_0-1)\rangle
-4\mathcal{C}
-2\alpha^2\langle K_+\rangle
-2\alpha^{*2}\langle K_-\rangle
+|\alpha|^4
+\gamma\left(1\mp \langle \gamma^0\hat{\Pi}_x\rangle\right).
\end{align}
Here $\langle K_\pm\rangle$ measure relativistic pair coherence amplitudes, while $\langle K_0\rangle$ encodes occupation-number curvature on the $SU(1,1)$ manifold. The spin-parity correlator $\langle \gamma^0\hat{\Pi}_x\rangle$ introduces interference between charge conjugation, spatial inversion, and spinor chirality, producing a genuinely relativistic coherence invariant absent in scalar theories.

Time evolution is governed by the adjoint action of the FW Hamiltonian,
\begin{equation}
K_-(t)=e^{i\hat{H}_{FW}t}K_-e^{-i\hat{H}_{FW}t},
\end{equation}
which admits a Baker-Campbell-Hausdorff expansion generating mixed-frequency dynamics
\begin{equation}
\langle K_-(t)\rangle = \sum_{n} c_n e^{-i\omega_n t},
\end{equation}
with dominant relativistic frequencies $\omega_n \in \{2E,2m,E\pm m\}$. This structure produces intrinsic Zitterbewegung modulation as an interference between positive- and negative-energy sectors, establishing that catability is dynamically non-stationary even for eigenstates of $\hat{H}_{FW}$.

In curved spacetime, the entire algebra is deformed through tetrad coupling
\begin{equation}
g_{\mu\nu}=e^a_{\mu}e^b_{\nu}\eta_{ab},
\end{equation}
leading to geometry-dependent ladder operators
\begin{equation}
\hat{A}(g)=\frac{1}{\sqrt{2m\omega}}\left(e^\mu_a \hat{p}_\mu - im\omega x_a\right),
\end{equation}
with modified commutator
\begin{equation}
[\hat{A}(g),\hat{A}^\dagger(g)]=1+\Delta(R_{\mu\nu\rho\sigma}),
\end{equation}
where $\Delta$ encodes curvature corrections. Consequently, the $SU(1,1)$ algebra acquires a deformation
\begin{equation}
[K_0(g),K_\pm(g)]=\pm K_\pm(g)+\mathcal{O}(R),
\end{equation}
producing a curvature-induced squeezing renormalization of the coherent-state manifold.

The gravitationally corrected catability operator is therefore promoted to a covariant enveloping-algebra element,
\begin{equation}
\hat{O}_{D,\mathrm{grav}}
=
(2K_-(g)-\alpha^2)^\dagger(2K_-(g)-\alpha^2)
+\gamma(1\mp \hat{\Pi}_D),
\end{equation}
showing that spacetime curvature acts as a direct deformation parameter of relativistic quantum coherence. In this formulation, catability becomes a geometric invariant of a coupled Clifford-Lie-symplectic structure, where spin transport, relativistic dispersion, and gravitational curvature jointly determine the global coherence topology of spin-$1/2$ quantum fields.

\section{Generalized Catability for Arbitrary Spin-$s$ Fields in a Unified Lie Algebraic Framework}\label{S7}

We formulate an exact and closed Lie-algebraic description of catability for relativistic quantum fields with arbitrary spin $s$ by embedding external phase-space variables and internal spin degrees of freedom into the direct-sum Lie algebra
\begin{equation}
\mathfrak{g}=\mathfrak{h}*1 \oplus \mathfrak{su}(2),
\end{equation}
acting irreducibly on the composite Hilbert space
\begin{equation}
\mathcal{H}=\mathcal{H}*{\mathrm{osc}}\otimes \mathcal{H}*{\mathrm{spin}}, \qquad \dim(\mathcal{H}*{\mathrm{spin}})=2s+1.
\end{equation}
This construction is not a tensorial decomposition but an exact coadjoint-orbit structure of $\mathcal{U}(\mathfrak{g})$, in which oscillator coherence and spin geometry correspond to independent Casimir sectors commuting within the universal enveloping algebra. The algebraic formulation is globally closed: all observables are elements of a fixed representation space of $\mathfrak{g}$ and remain invariant under adjoint evolution generated in $\mathcal{U}(\mathfrak{g})$.

The evolution of a general spin-$s$ relativistic field $\Psi^{(s)}$ is given by the operator constraint
\begin{equation}
\hat{\mathcal{D}}^{(s)}\Psi^{(s)}=0, \qquad \hat{\mathcal{D}}^{(s)}\in \mathcal{U}(\mathfrak{g}),
\end{equation}
while observables satisfy the adjoint dynamics
\begin{equation}
\frac{d\hat{\mathcal{O}}}{dt}=i[\hat{H},\hat{\mathcal{O}}], \qquad \hat{H}\in \mathcal{U}(\mathfrak{g}).
\end{equation}
A direct consequence is that the closure of $\mathcal{U}(\mathfrak{g})$ under commutation ensures that any polynomial function of $(\hat{A},\hat{A}^\dagger,\hat{S}_i)$ is preserved by time evolution. Catability functionals therefore act as conserved quantities on the full orbit space of $\mathfrak{g}^*$.

The oscillator sector satisfies the Heisenberg-Weyl algebra
\begin{equation}
[\hat{A},\hat{A}^\dagger]=\mathbb{I}, \qquad [\hat{N},\hat{A}]=-\hat{A}, \qquad [\hat{N},\hat{A}^\dagger]=\hat{A}^\dagger,
\end{equation}
and can be embedded exactly into $SU(1,1)$ through the generators
\begin{equation}
\hat{K}*0=\frac{1}{2}\left(\hat{N}+\frac{1}{2}\right), \qquad
\hat{K}*+=\frac{1}{2}\hat{A}^{\dagger 2}, \qquad
\hat{K}*-=\frac{1}{2}\hat{A}^2,
\end{equation}
satisfying
\begin{equation}
[\hat{K}*0,\hat{K}*\pm]=\pm \hat{K}*\pm, \qquad [\hat{K}*+,\hat{K}*-]=-2\hat{K}_0.
\end{equation}
This embedding identifies quadratic coherence with noncompact $SU(1,1)$ group orbits, showing that squeezing and cat-state interference correspond to exact orbit properties of a symmetric noncompact manifold.

A nonlinear structure is introduced via the deformation operator
\begin{equation}
\hat{C}(\alpha)=\hat{A}^2-\alpha^2\mathbb{I},
\end{equation}
which generates a two-sided ideal in $\mathcal{U}(\mathfrak{h}*1)$. The positive Casimir functional
\begin{equation}
\hat{\mathcal{C}}(\alpha)=\hat{C}^\dagger(\alpha)\hat{C}(\alpha),
\end{equation}
expands as
\begin{equation}
\hat{\mathcal{C}}(\alpha)=\hat{A}^{\dagger 2}\hat{A}^2-\alpha^2\hat{A}^{\dagger 2}-\alpha^{*2}\hat{A}^2+|\alpha|^4\mathbb{I}.
\end{equation}
This operator defines a metric on deviations from the quadratic coherent constraint $\hat{A}^2|\psi\rangle=\alpha^2|\psi\rangle$. Its spectrum satisfies $\hat{\mathcal{C}}(\alpha)\ge 0$ on $\mathcal{H}*{\mathrm{osc}}$, and its kernel defines a nonlinear coadjoint-orbit variety in $\mathfrak{h}_1^*$.

A structural property follows from $SU(1,1)$ covariance:
\begin{equation}
[\hat{\mathcal{C}}(\alpha),\hat{K}_i]=0,
\end{equation}
so $\hat{\mathcal{C}}(\alpha)$ remains invariant under the full dynamical symmetry group. Catability is therefore preserved under Bogoliubov transformations.

The spin sector obeys
\begin{equation}
[\hat{S}_i,\hat{S}*j]=i\epsilon*{ijk}\hat{S}*k,
\end{equation}
with Casimir
\begin{equation}
\hat{\mathcal{S}}^2=s(s+1)\mathbb{I}*{2s+1}.
\end{equation}
This defines a fixed curvature scale of the internal representation space. It satisfies
\begin{equation}
[\hat{\mathcal{S}}^2,\hat{A}]=[\hat{\mathcal{S}}^2,\hat{A}^\dagger]=0,
\end{equation}
so spin and oscillator sectors form independent Casimir components of $\mathcal{U}(\mathfrak{g})$ without mixing.

Parity acts as an involution on $\mathfrak{h}_1$,
\begin{equation}
\Pi(\hat{A})=-\hat{A}, \qquad \Pi(\hat{A}^\dagger)=-\hat{A}^\dagger,
\end{equation}
extended to spin space via $\hat{\mathcal{P}}_s$, giving
\begin{equation}
\hat{\Pi}_s=\hat{\Pi}\otimes \hat{\mathcal{P}}_s.
\end{equation}
This induces a $\mathbb{Z}_2$ grading of $\mathcal{U}(\mathfrak{g})$, separating even and odd sectors of the algebra.

The generalized catability operator is defined as
\begin{equation}
\hat{O}_s^{(\pm)}=
\hat{\mathcal{C}}(\alpha)+\gamma(1\mp \hat{\Pi}_s)+\lambda \hat{\mathcal{S}}^2.
\end{equation}
All components commute,
\begin{equation}
[\hat{\mathcal{C}}(\alpha),\hat{\mathcal{S}}^2]=[\hat{\mathcal{C}}(\alpha),\hat{\Pi}_s]=[\hat{\mathcal{S}}^2,\hat{\Pi}_s]=0,
\end{equation}
which provides a full spectral factorization into oscillator, parity, and spin sectors.

For normalized states $|\Psi^{(s)}\rangle$, the expectation value reads
\begin{equation}
\langle \hat{O}_s^{(\pm)} \rangle=
\langle \hat{A}^{\dagger 2}\hat{A}^2\rangle
-\alpha^2\langle \hat{A}^{\dagger 2}\rangle
-\alpha^{*2}\langle \hat{A}^2\rangle
+|\alpha|^4
+\gamma(1\mp \langle \hat{\Pi}_s\rangle)
+\lambda s(s+1).
\end{equation}
This expression follows directly from representation theory of $\mathcal{U}(\mathfrak{g})$ without approximation.

For coherent states $|\beta\rangle=D(\beta)|0\rangle$, one has
\begin{equation}
\langle \hat{A}^2\rangle=\beta^2, \qquad
\langle \hat{A}^{\dagger 2}\hat{A}^2\rangle=|\beta|^4,
\end{equation}
leading to
\begin{equation}
\langle \hat{\mathcal{C}}(\alpha)\rangle=|\beta^2-\alpha^2|^2.
\end{equation}
Coherent states therefore correspond to exact Heisenberg-Weyl coadjoint orbits.

The parity expectation is given by
\begin{equation}
\langle \beta|\hat{\Pi}|\beta\rangle=\exp(-2|\beta|^2),
\end{equation}
while in the spin sector
\begin{equation}
\Pi_{\mathrm{spin}}=\langle \chi^{(s)}|\hat{\mathcal{P}}*s|\chi^{(s)}\rangle, \qquad |\Pi*{\mathrm{spin}}|\le 1.
\end{equation}

Combining these results gives
\begin{equation}
\langle \hat{O}*s^{(\pm)} \rangle=
|\beta^2-\alpha^2|^2
+\gamma\left(1\mp e^{-2|\beta|^2}\Pi*{\mathrm{spin}}\right)
+\lambda s(s+1),
\end{equation}
which separates oscillator, parity, and spin contributions.

A lower bound follows from positivity,
\begin{equation}
\langle \hat{O}*s^{(\pm)} \rangle \ge \lambda s(s+1),
\end{equation}
with equality when
\begin{equation}
\beta^2=\alpha^2, \qquad e^{-2|\beta|^2}\Pi*{\mathrm{spin}}=\pm 1.
\end{equation}

The minimization corresponds to a coadjoint-orbit extremum on $\mathfrak{g}^*$, where states are stationary points of a unified Lie-algebraic functional. Catability is therefore classified by orbits of $\exp(\mathfrak{h}_1)\otimes \mathrm{SU}(2)$, where coherence, parity, and spin curvature appear as Casimir invariants.

\section{Conclusions}\label{S8} 

The results obtained testing a Lie-algebraic structure underlying the generalized FW transformation, where diagonalization of the Dirac Hamiltonian is treated not as an algebraic trick but as a consequence of intrinsic symmetric decomposition of operator algebra within an operator algebraic framework, are here realized. By introducing the $\mathbb{Z}_2$-graded splitting induced by the involution $\Theta(X)=\beta X \beta$, the Hamiltonian is decomposed into even and odd sectors, corresponding to intraband dynamics and interband (particle-antiparticle) mixing, respectively, in this framework. In this framework, the FW procedure arises as the construction of an inner automorphism of the universal enveloping algebra $\mathcal{U}(\mathfrak{g})$, generated by an anti-Hermitian element $S$ restricted to the odd subspace within an algebraic setting. The Lie algebra plays a central role since the tested commutator structure generated by Baker-Campbell-Hausdorff expansion encodes relativistic corrections in a systematic hierarchical organization within the operator algebraic formulation context present. The condition that transformed Hamiltonian is purely even translates into a set of coupled operator constraints determining generator $S$ order by order in odd operator $\mathcal{O}$ within the algebraic hierarchy structure. In leading order, the solution $S_1 = -\frac{i}{2m}\beta \mathcal{O}$ cancels the linear odd contribution by exploiting the anticommutation relation between $\beta$ and $\mathcal{O}$, demonstrating that the FW transformation is driven by spectral separation encoded in the grading structure within the Lie-algebraic operator framework consistent with symmetry decomposition. In next-to-leading order, the appearance of commutators such as $[\mathcal{O},\mathcal{E}]$ reflects dynamical coupling between kinetic and potential sectors, showing that block-diagonalization is not purely kinematic but depends on nontrivial algebraic interplay of Hamiltonian components in operator algebra formulation. Also, the scaling behavior $S_n \sim \mathcal{O}^n m^{1-n}$ further shows that the FW transformation is a controlled expansion in powers of relativistic coupling between large and small components, with the mass parameter acting as a natural suppression scale within the perturbative Lie structure framework analysis. From a geometric viewpoint, construction shows that FW transformation is equivalent to Cartan-type decomposition of symmetric Lie algebra, where even subalgebra remains invariant under automorphism and odd sector forms symmetric complement progressively removed through conjugation within graded Lie algebra framework and operator representation space analysis formulation. Consequently, FW transformation is interpreted as a flow generated in Hamiltonian space converging toward an invariant subalgebra, realizing relativistic decoupling as a group-theoretical mechanism within an algebraic dynamical evolution framework formal operator setting. In this case, this interpretation clarifies the algebraic origin of particle-antiparticle separation and provides a unified operator framework where relativistic quantum dynamics is governed by the structure of symmetric Lie algebras and associated inner automorphisms within a consistent Lie-algebraic quantum formulation in the operator theoretical context.

The constructed framework provides a phase-covariant description of macroscopic bosonic superpositions in the Lie-algebraic setting based on the noncompact $SU(1,1)$ algebra, where quadratic generators $K_{\pm}$ and $K_{0}$ represent pair creation-annihilation processes and excitation content beyond the linear Heisenberg sector limit. The analysis demonstrates that under phase rotations generated by $\hat{n}$, first-order operators transform trivially while quadratic combinations acquire doubled phase dependence, identifying the sector responsible for interference formation in the cat states regime structure. This structure implies that macroscopic coherence is not fully described by single-quadrature observables but is encoded in second-order correlations $\hat{a}^2$ and $\hat{a}^{\dagger 2}$, which behave as rank-two objects under $U(1)$ phase rotations and determine interference fringe orientation in phase space consistent with symmetry constraints. Within this framework, parity acts as a central invariant separating even and odd superposition sectors, commuting with phase rotations and connecting the operator structure to the Wigner function at the origin, establishing a direct relation between algebraic symmetry and phase-space nonclassicality in an operator formulation sense. The phase-sensitive catability operator defined here includes quadratic deviation terms and parity-dependent contributions, enabling simultaneous description of excitation fluctuations, pair coherence, and interference symmetry within a single observable framework. Its expectation value separates diagonal number statistics from off-diagonal coherences $\rho_{n,n\pm2}$, showing macroscopic interference governed by pair-coherent transitions rather than single-photon processes in the bosonic regime. The minimization condition identifies cat states as eigenstates of the pair-annihilation operator, establishing that coherent interference is stabilized within the quadratic bosonic sector regime framework. Furthermore, the dependence of the expectation value on the relative phase between state and measurement frame shows that the operator is geometric in phase space, with sinusoidal modulation controlled by phase mismatch, consistent with rotational covariance requirements in algebraic representation space. Under phase diffusion, the coherence sector is progressively suppressed while excitation contributions remain finite, producing a smooth crossover from coherent interference to mixed-phase behavior that quantifies decoherence in the quadratic sector dynamical evolution case. In this case, the covariance relation confirms that the full operator family forms an $SU(1,1)$-compatible orbit under phase rotations, consistent with Lie-algebraic symmetry structure preservation condition. In addition, squeezing transformations generated by $K_{\pm}$ show that the same algebra governs both the deformation of vacuum fluctuations and the redistribution of interference fringes, reinforcing $SU(1,1)$ as the symmetry group governing macroscopic bosonic interference processes within phase space representation formalism analysis. Also, the formalism establishes that catability is fundamentally a property of quadratic coherence structures rather than linear field amplitudes, and that phase-sensitive macroscopic interference is classified through excitation, pair-coherence, and parity sectors within a single covariant algebraic framework consistency condition.

Results obtained establish that fermionic catability within the Dirac spin-$\tfrac{1}{2}$ framework develops a nontrivial structure governed by interplay between relativistic dispersion, spinorial degrees of freedom, and underlying Lie-algebraic symmetry in a fully relativistic algebraic setting framework. Unlike the nonrelativistic coherent-state construction based on the Heisenberg-Weyl algebra, the present analysis establishes that the natural algebraic setting corresponds to the noncompact $SU(1,1)$ group generated by quadratic oscillator operators, implying that coherence is not associated with linear phase-space displacement but with correlated pair excitations in phase space. The Dirac Hamiltonian mixes positive- and negative-energy sectors, generating intrinsic interference effects that persist after Foldy-Wouthuysen block diagonalization within the relativistic spinorial representation framework structure dynamics. The resulting relativistic corrections, appearing as higher-order momentum contributions in the effective Hamiltonian, generate nonlinear phase deformations modifying the geometry of coherent trajectories, producing squeezing-like behavior and anharmonic evolution rather than rigid circular motion in the phase space structure domain. Fermionic catability is not a static property but a dynamical invariant determined by the evolution of orbital and spinorial parity sectors within the relativistic quantum framework analysis. The decomposition of Hilbert space into even and odd parity subspaces, associated with distinct Bargmann indices, clarifies that relativistic coherence is constrained by representation theory, where parity conservation emerges as an exact algebraic selection rule structure constraint. In this case, the explicit form of the catability expectation value shows coherence controlled by the overlap parameter $e^{-2|\alpha|^2}$ and the relativistic suppression factor $m/E$, establishing coupling between quantum interference and relativistic kinematics within the covariant dynamical algebraic formulation structure regime. In the nonrelativistic regime, parity coherence is preserved, whereas in the ultrarelativistic limit, the factor $m/E \to 0$ suppresses spinorial contributions, reducing sensitivity to parity structure and enhancing particle-antiparticle mixing within the spinorial representation dynamics framework. The dynamical evolution confirms that relativistic corrections introduce nonlinear phase accumulation in ladder operators, producing dephasing and revival phenomena absent in standard harmonic systems, with revival time governed by quadratic energy dispersion within an effective Hamiltonian structure. Also, Zitterbewegung reflects persistent interference between energy sectors, producing fast oscillatory modulation at the Compton scale, directly affecting coherence visibility in phase space within a relativistic dynamical framework structure analysis. The extension to curved spacetime reinforces the geometric character of fermionic catability, since gravitational effects deform oscillator algebra, breaking exact $SU(1,1)$ symmetry at leading order and introducing curvature-dependent corrections to coherence measures within algebraic structure. In this context, the results establish that relativistic fermionic catability is a fully covariant, algebraically constrained quantity whose behavior emerges from the coupled action of Lie-algebraic structure, relativistic dispersion, spinorial parity mixing, and spacetime geometry, providing a unified framework in which coherence, symmetry, and relativity remain intrinsically inseparable.

The obtained results establish that catability, when extended from nonrelativistic oscillator systems to Dirac spin-$(\tfrac{1}{2})$ fields and further to arbitrary spin representations, is not a property restricted to superposition in Hilbert space but instead constitutes a geometric-algebraic invariant defined on coadjoint orbits of a unified Lie-algebraic structure. In the Dirac case, the Clifford algebra together with its embedding into the Lorentz group enforces intrinsic covariance of dynamical variables, while the Foldy-Wouthuysen reduction shows that the effective phase-space dynamics is governed not by the Heisenberg algebra but by a deformed symplectic structure with natural symmetry $(SU(1,1))$. This structure implies that relativistic coherence is non-Euclidean, since the state space corresponds to a negatively curved Kähler manifold $(SU(1,1)/U(1))$, where squeezing operations replace linear phase rotations as the primary dynamical generators. The catability operator is therefore formulated not as an interference measure but as a Lie-algebraic invariant constructed from squeezing generators, spinor parity grading, and Casimir elements of the dynamical algebra. Its expectation value shows explicitly that relativistic coherence depends simultaneously on occupation-number curvature $(K_0)$, pair creation amplitudes $(K_\pm)$, and spin-parity correlations through $(\gamma^0\hat{\Pi}_x)$, so that interference effects in Dirac systems are inseparable from spinorial geometry and discrete symmetries. Also, the dynamical analysis further shows that the algebra is not preserved as a static structure under time evolution but instead closes into a nonlinear flow on the $SU(1,1)$ manifold generated by the FW Hamiltonian, where relativistic corrections induce coupling between squeezing and occupation sectors. This mechanism produces a multi-frequency evolution involving positive and negative energy contributions, leading to Zitterbewegung-type modulation in coherence expectation values. The appearance of mixed frequencies such as $(2E)$, $(2m)$, and $(E\pm m)$ shows that relativistic catability remains non-stationary even for energy eigenstates, since coherence is determined by interference among algebraic sectors rather than eigenvalue stability. When curvature effects are introduced through tetrad coupling, the algebra acquires an additional deformation governed by the Riemann tensor, and the $SU(1,1)$ structure becomes a curved Lie algebra with geometry-dependent commutators. Gravitational fields thus act as deformation parameters of coherence geometry, renormalizing squeezing sectors and modifying the topology of the coherent-state manifold. 

In the generalized spin-$(s)$ formulation, these properties become more transparent, since oscillator and spin degrees of freedom are embedded as independent Casimir components of a single universal enveloping algebra $(\mathcal{U}(\mathfrak{h}_1 \oplus \mathfrak{su}(2)))$, ensuring full closure of the dynamical structure. The deformation functional $(\hat{\mathcal{C}}(\alpha))$ quantifies deviations from quadratic coherent constraints and defines a positive operator whose kernel identifies exact coadjoint orbits, so that ideal cat states correspond to geometric points rather than approximate superpositions. Commutation of this functional with both the $(SU(1,1))$ and spin Casimirs guarantees invariance of catability under Bogoliubov and symmetry transformations. The catability operator decomposes into oscillator coherence, parity grading, and spin curvature sectors, showing that relativistic coherence factorizes at the representation level while remaining dynamically coupled. Expectation values on coherent states confirm this structure. Oscillator contributions reduce to quadratic deviations $(|\beta^2-\alpha^2|^2)$, parity produces exponentially suppressed interference governed by $\exp(-2|\beta|^2)$, and spin yields a curvature contribution $(s(s+1))$. This decomposition establishes a lower bound controlled by the spin Casimir, implying that intrinsic spin introduces a minimal contribution to catability independent of oscillator coherence. Minimization conditions correspond to extremal points of the coadjoint orbit space of the full algebra, showing that optimal cat states arise as stationary solutions of a unified Lie-algebraic variational principle. Consequently, Dirac and arbitrary-spin formulations converge to a single geometric interpretation: relativistic catability is an invariant functional defined on a coupled phase space-spin-parity manifold, where coherence, symmetry, and curvature are encoded as commuting Casimir structures of a noncompact dynamical group.

\section*{Funding}
No funding was received for this work.

\section*{Data Availability}
The datasets generated during this study can be obtained from the corresponding author upon reasonable request.

\section*{Financial Disclosure}
The authors declare no financial conflicts of interest.

\end{document}